\documentclass[twocolumn,prl,showpacs]{revtex4-1}

\usepackage{dcolumn}
\usepackage{amsfonts}
\usepackage{amsmath}
\usepackage{amssymb}
\usepackage{bm}
\usepackage{epsfig}
\usepackage[usenames]{color}

\begin{document}

\newcommand{\brm}[1]{\bm{{\rm #1}}}
\newcommand{\tens}[1]{\underline{\underline{#1}}}
\newcommand{\xv}{\bm{{\rm x}}}
\newcommand{\Rv}{\bm{{\rm R}}}
\newcommand{\uv}{\bm{{\rm u}}}
\newcommand{\nv}{\bm{{\rm n}}}
\newcommand{\Nv}{\bm{{\rm N}}}
\newcommand{\Mv}{\bm{{\rm M}}}
\newcommand{\Av}{\bm{{\rm A}}}
\newcommand{\ev}{\bm{{\rm e}}}
\newcommand{\Cv}{\bm{{\rm C}}}
\newcommand{\Qv}{\bm{{\rm Q}}}
\newcommand{\Sv}{\bm{{\rm S}}}
\newcommand{\qv}{\bm{{\rm q}}}
\newcommand{\fv}{\bm{{\rm f}}}
\newcommand{\tv}{\bm{{\rm t}}}
\newcommand{\kv}{\bm{{\rm k}}}
\newcommand{\Dv}{\bm{{\rm D}}}
\newcommand{\Hv}{\bm{{\rm H}}}
\newcommand{\piv}{\bm{{\rm \pi}}}
\newcommand{\Piv}{\bm{{\rm \Pi}}}
\newcommand{\Wv}{\bm{{\rm W}}}
\newcommand{\Zv}{\bm{{\rm Z}}}
\newcommand{\Bv}{\bm{{\rm B}}}
\newcommand{\av}{\bm{{\rm a}}}
\newcommand{\bv}{\bm{{\rm b}}}
\newcommand{\Gv}{\bm{{\rm G}}}
\newcommand{\Tv}{\bm{{\rm T}}}
\newcommand{\pv}{\bm{{\rm p}}}
\newcommand{\wv}{\bm{{\rm w}}}
\newcommand{\wn}{m}
\newcommand{\rv}{\bm{{\rm r}}}
\newcommand{\cvh}{\hat{\brm{c}}}
\newcommand{\evh}{\hat{\brm{e}}}
\newcommand{\nvh}{\hat{\brm{n}}}
\newcommand{\avh}{\hat{\brm{a}}}
\newcommand{\bvh}{\hat{\brm{b}}}
\newcommand{\re}{\text{Re}}
\newcommand{\Ochange}[1]{{\color{red}{#1}}}
\newcommand{\Ocomment}[1]{{\color{green}{#1}}}
\newcommand{\Tcomment}[1]{{\color{blue}{#1}}}
\newcommand{\Tchange}[1]{{\color{red}{#1}}}
\newcommand{\Remove}[1]{{\color{yellow}{#1}}}

\title{Signatures of Topological Phonons in Superisostatic Lattices}

\author{Olaf Stenull}
\affiliation{Department of Physics and Astronomy, University of
Pennsylvania, Philadelphia, PA 19104, USA }

\author{T. C. Lubensky}
\affiliation{Department of Physics and Astronomy, University of
Pennsylvania, Philadelphia, PA 19104, USA }

\vspace{10mm}
\date{\today}

\begin{abstract}
Soft topological surface phonons in idealized ball-and-spring lattices with coordination number $z=2d$ in $d$ dimensions become finite-frequency surface phonons in physically realizable superisostatic lattices with $z>2d$. We study these finite-frequency modes in model lattices with added next-nearest-neighbor springs or bending forces at nodes with an eye to signatures of the topological surface modes that are retained in the physical lattices. Our results apply to metamaterial lattices, prepared with modern printing techniques, that closely approach isostaticity.
\end{abstract}


\maketitle
Recent work \cite{Kane2014,Lubensky2015,MaoLub2018} laid the foundation for a theory, akin to the topological band theory of electronic materials such as quantum Hall systems \cite{Halperin1982,Haldane1983} and topological insulators \cite{KaneMele2005a,KaneMele2005b,BernevigZha2006,MooreBal2007,FuMele2007,Hasan2010,QiZhang2011}, of topological mechanics of periodic ball-and-spring isostatic lattices with average coordination number $z$, under periodic boundary conditions, equal to twice the spatial dimension, $2d$. This theory predicts the existence of zero-energy surface-modes at every surface wavenumber with the number of these modes on different surfaces depending on the topological properties of the bulk phonon  spectrum. It has been applied to a variety of systems and phenomena \cite{Stenull2014,Paulose2015,Paulose2015a,SussmanLub2016,Rocklin2016,StenullLub2016,ChenSan2016,MeeussenVit2016,BaardinkVit2018,ZhouMao2018,ZhouMao2018b} from random and jammed systems to stress concentration at topological domain walls.
Our focus here is on periodic fully gapped systems in which the only bulk zero modes are those imposed by translational invariance at wavenumber $\qv=\brm{0}$.  Naturally  occurring crystals always have an effective coordination number  greater than $2d$ (because forces between sites have a range greater than the inter-site separation) or stabilizing bending forces favoring particular angles between bonds incident on a given site, and they are not candidates to exhibit topological mechanics.  On the other hand, with the aid of modern printing and cutting techniques, metamaterials with $z=2d$ consisting of vertices connected by thin nearest-neighbor (NN) elastic beams can be designed \cite{StenullLub2016,BaardinkVit2018} and constructed \cite{BilalHuber2017, MaZhou2018} to minimize bending forces and, thereby, closely approach the isostatic limit to which the topological theory of Refs.~\cite{Kane2014,Lubensky2015,MaoLub2018} applies. 

\begin{figure*}[ptb]
	\includegraphics[width=4.9cm]{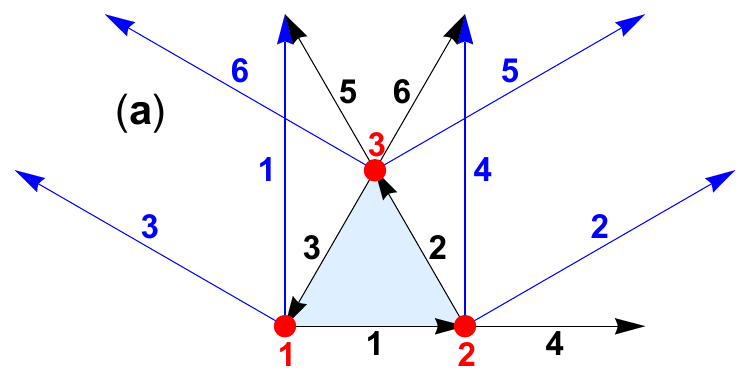}
	\includegraphics[width=3.9cm]{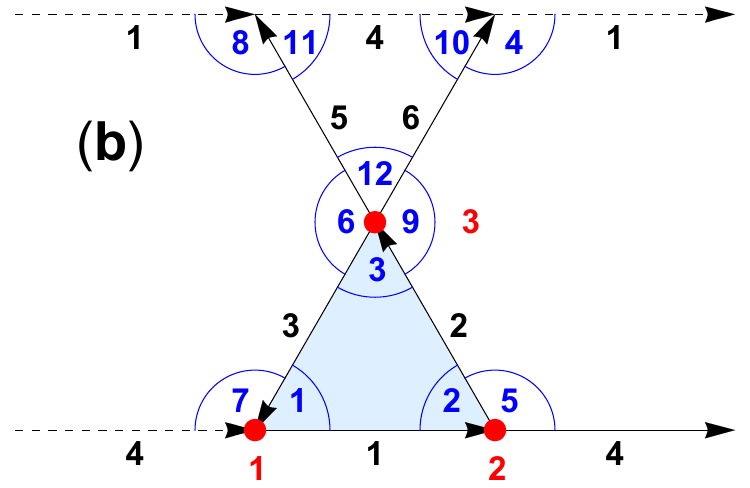}
	\hspace{0.3cm}
	\includegraphics[width=3.6cm]{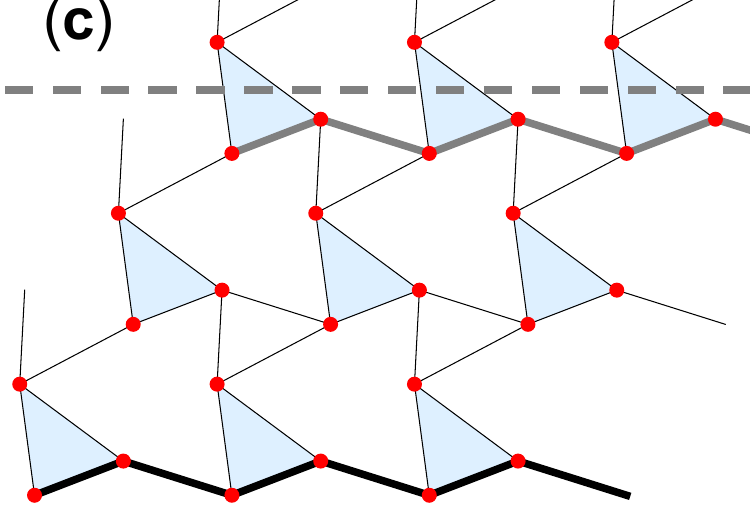}
	\hspace{0.3cm}
	\includegraphics[width=4.4cm]{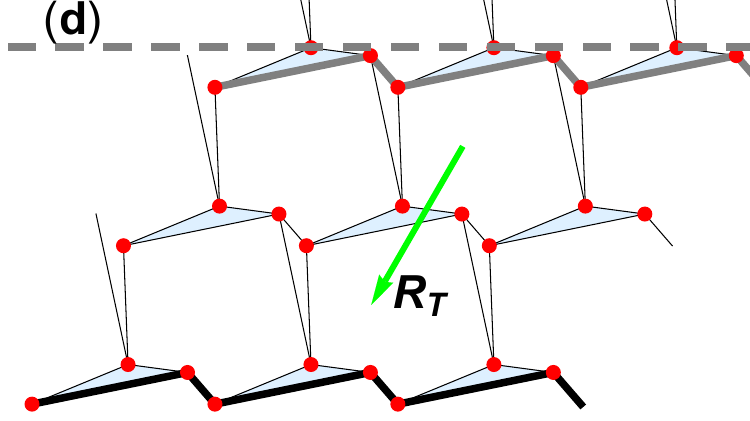}
	\caption{(a) Unit cell of the KL with NN bonds (black) and additional NNN bonds (blue). (b) Unit cell of the KL with bending energies (blue arcs). (c) The $X_{\text{nt}}$ and (d) the $X_{\text{t}}$ conformation. The thick black bonds mark our bottom surface, and the thick dashed line marks a possible cut to liberate a top surface. The green arrow indicates the topological polarization $R_T$ of $X_{\text{t}}$.}%
	\label{fig:genKagomeNNN}%
\end{figure*}

Here we study generalized kagome lattices (GKLs)  to which weak next-nearest-neighbor (NNN) springs or bending forces~\cite{Mao2011a} are added [Fig.~\ref{fig:genKagomeNNN}], and we focus on how their surface modes evolve as the magnitudes $v$ of these forces are increased from zero.  In the presence of either such force, the originally isostatic lattices become stable elastic materials whose long-wavelength excitations are described by continuum elasticity, which predicts identical Rayleigh waves \cite{Landau-elasticity} on opposite surfaces of a strip [see Supplemental Material (SM)]. It would be natural to expect that these Rayleigh waves evolve from zero-energy surface modes of the isostatic lattice, and this is indeed the case for non-topological lattices, which   have the same number of zero modes on all pairs  of opposite parallel surfaces \cite{Kane2014}. But this cannot be the case for topological lattices, which have opposite parallel surfaces with different numbers of zero modes --at the extreme no zero modes on one and an associated excess of zero modes on the opposite surface. In what follows, we discuss Rayleigh waves in weakly superisostatic lattices in the context of topological phonons, and we detail how the dilemma posed by the topological lattices is resolved. 

For the sake of generality, we consider  generic non-topological ($X_{\text{nt}}$) and topological ($X_{\text{t}}$) GKLs that have the lowest possible plane crystallographic symmetry, p1. To study surface modes, we assume that a free surface parallel to the $x$-axis exists as indicated in Fig.~\ref{fig:genKagomeNNN} so that the network as a whole is semi-infinite with a parallel opposite surface located at infinite distance. For the standard GKL with $v=0$, liberating these two surfaces from the constraints of periodic boundary conditions amounts to removing 2 bonds or 4 bonds and one site per surface unit cell. Both choices lead to two zero-surface-modes per surface wavenumber $q$ distributed on the combined lower and upper surfaces, but the latter, which we consider, has smoother upper and lower surfaces as shown in Fig.~\ref{fig:genKagomeNNN}. The topological polarization $\Rv_T$~\cite{Kane2014} calculated from the bulk phonon spectrum is zero for $X_{\text{nt}}$, and it is non-zero and pointing towards the bottom surface, $\Rv_T = -\frac{1}{2} (1, \sqrt{3})$, for $X_{\text{t}}$. As a consequence, there is one zero-surface-mode per $q$ on either surface for $X_{\text{nt}}$ and two (zero) zero-surface-modes per $q$ on the bottom (top) surface for $X_{\text{t}}$.

\begin{figure*}[ptb]
\begin{minipage}{5.5cm}
	\includegraphics[width=5.3cm]{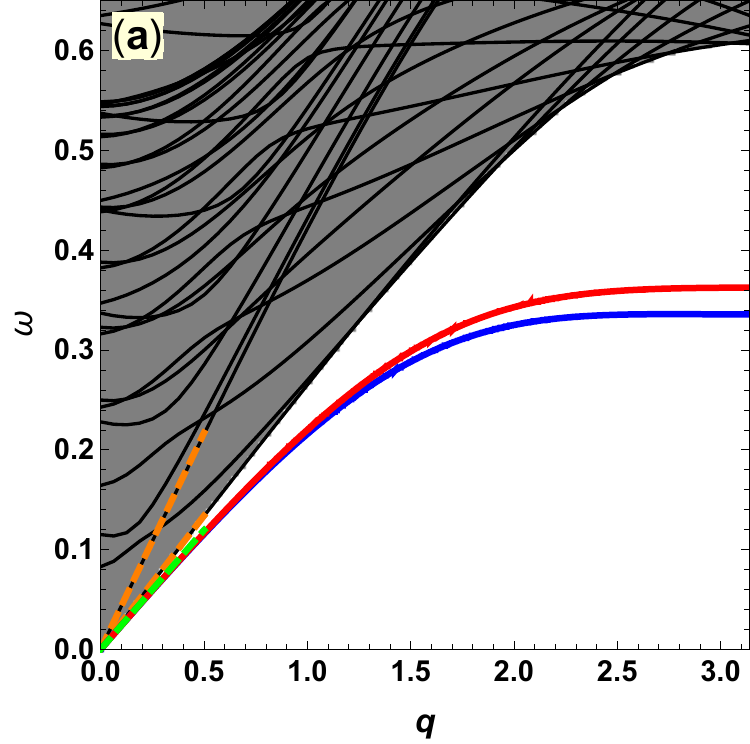}
	\includegraphics[width=5.3cm]{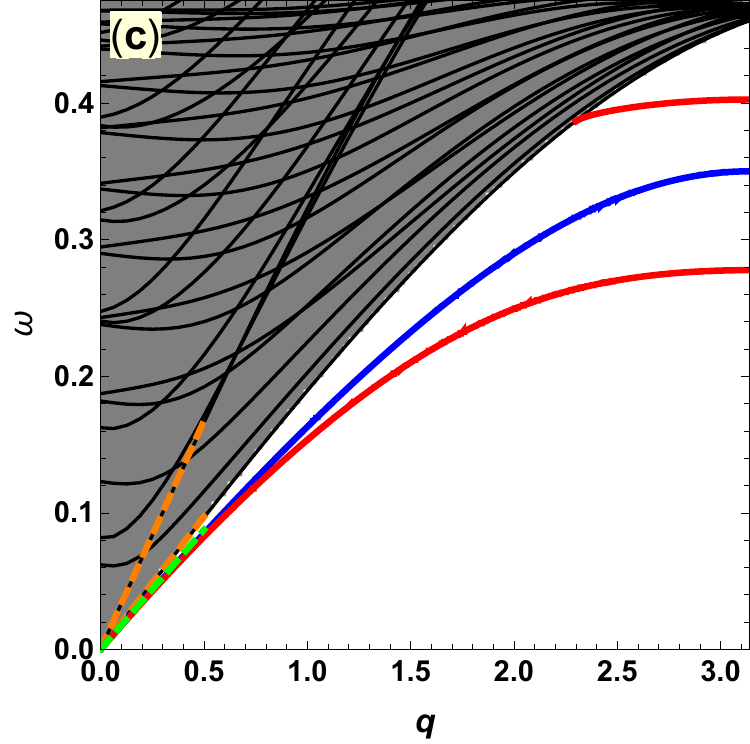}
	\end{minipage}
\begin{minipage}{5.5cm}
	\includegraphics[width=5.3cm]{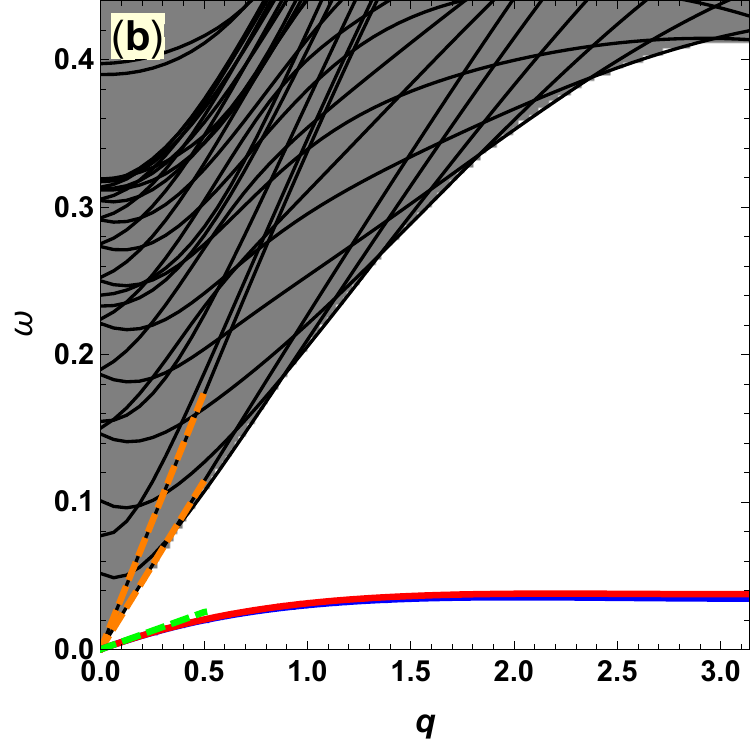}
	\includegraphics[width=5.3cm]{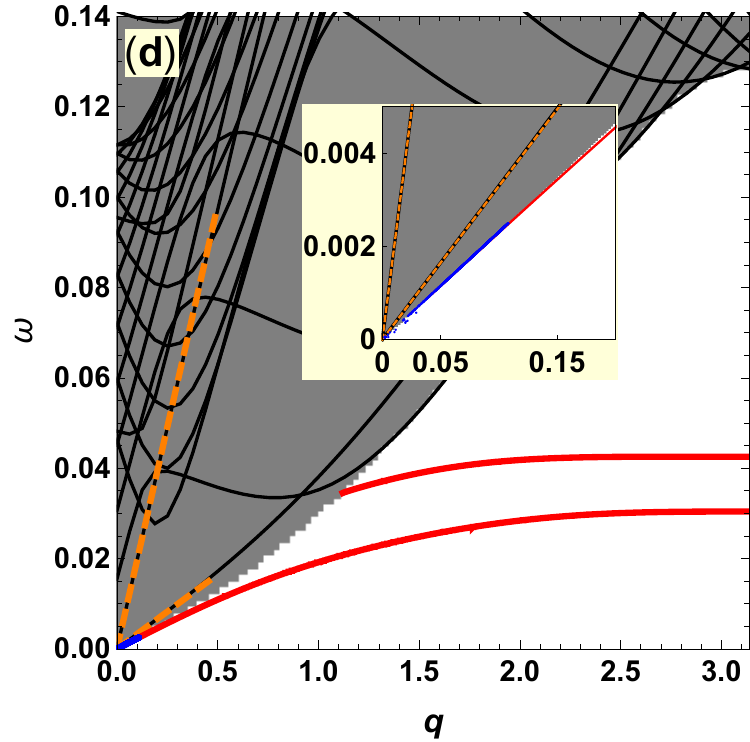}
\end{minipage}
\begin{minipage}{6.5cm}
	 \includegraphics[width=5.7cm]{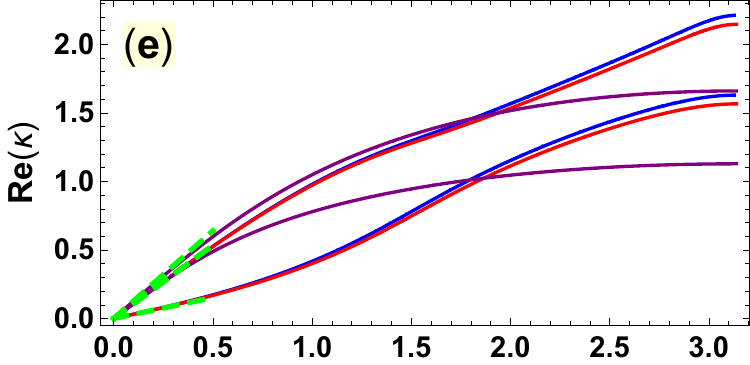}
	\includegraphics[width=5.7cm]{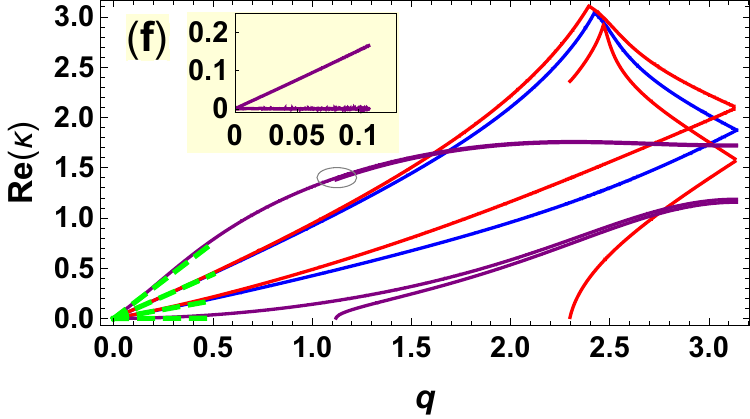}\\
	\vspace{2.5mm}
	\includegraphics[width=6.0cm]{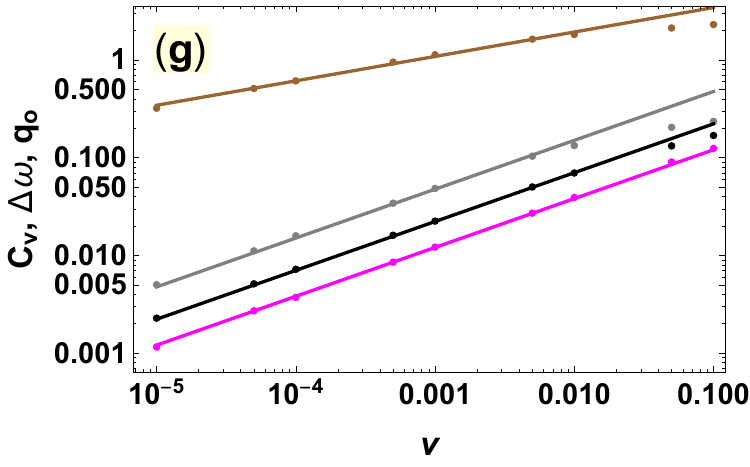}
\end{minipage}
\caption{Low-frequency mode structure for (a) $X_{\text{nt}}$, $v=0.1$; (b) $X_{\text{nt}}$, $v=0.001$; (c) $X_{\text{t}}$, $v=0.1$; (d) $X_{\text{t}}$, $v=0.001$. The gray areas are the projected bulk bands, and the black curves within these bands are bulk mode frequencies as a function of $q = q_x$ for different values of $q_y$, $q_y = 0, \pi/10, \pi/5, \ldots$. Note the strongly non-monotonic behavior of these modes for $X_{\text{t}}$ at $v=0.001$, a consequence of the lobes in density plots of the lowest mode at $v=0$ with $\omega \sim (q_y^2 -\beta q_x^2)$ with $\beta>0$ determined by $X_{\text{t}}$ \cite{Kane2014,Lubensky2015}. The color codes for curves in (a)-(d) are red –- bottom surface, blue –- top surface, orange –- longitudinal and transverse bulk sound modes at $q_y=0$, and green -– Rayleigh waves predicted by elasticity theory. Each surface mode is a linear combination of four modes that decay with $y$. The two smallest $\kappa$ (largest penetration depths) are plotted in (e) [(f)] for each surface mode shown in (a) and (b) [(c) and (d)]. The color codes for these curves are red (blue) for bottom (top) surface modes  at $v=0.1$, purple for bottom and top acoustic modes and for the optical mode at $v=0.001$, and dashed green for elastic theory results. Each of the purple curves in (e) consists of a nearly degenerate pair. The gray ellipsis in (f) highlights a hard-to-see onset of the acoustic mode after which the upper purple curve consists of a nearly degenerate pair. (g) $c_R$ (gray) for $X_{\text{nt}}$ and $c_R$ (black), $\Delta \omega$ (magenta), and $q_0$ (brown) for $X_{\text{t}}$. The lines in the corresponding colors are power-law fits with $c_R\sim\Delta \omega\sim v^{0.5}$ and  $q_0 \sim v^{0.25}$, in agreement with our crude estimates. Note that the surface modes for $v=0.001$ remain small and nearly flat throughout the SBZ implying low-energy point-like surface invaginations such as observed in Ref.~\cite{BilalHuber2017}.}%
	\label{fig:bands}%
\end{figure*}

Turning to $v>0$, we will first review our results and then present some details about how we obtained them. Because of space constraints and for concreteness, we center our discussion on the case with NNN forces. Further details, model elastic energies etc.,  and results for the case with bending forces are provided in the SM. Figure \ref{fig:bands} summarizes our major results about changes in the phonon band structure as the strength of the  NNN coupling increases from zero and, in particular, how long-wavelength  Rayleigh  waves with the same speed develop on opposite surfaces and how the zero-energy surface states at $v=0$ evolve with increasing  $v$. At $v=0.1$, both $X_{\text{nt}}$ and $X_\text{t}$ have one acoustic surface mode on each surface at each wavenumber $q$ in the surface Brillouin zone (SBZ). At small $q$, the  modes reduce to the elastic Rayleigh waves with dispersion $\omega_{R}(q) = c_R q$ on opposite surfaces with the same surface $c_R$, predicted by elastic theory. The situation at $v=0.001$ is very similar to that at $v=0.1$ for $X_{\text{nt}}$ except that the acoustic surface-mode frequency $\omega_s (q)$ is smaller at  every $q$, indicating an approach to a single zero mode at each $q$ on each surface as $v\rightarrow 0$. Figure \ref{fig:bands} (g) shows that $c_R\propto \sqrt{v}$ as follows from the observation that $\omega_s^2(q)$ must be linearly proportional to some combination of  spring constants and be equal to zero at $v=0$. The situation for $X_{\text{t}}$ is more complex.  The bottom surface has an acoustic mode that stretches across the SBZ and reduces to the expected Rayleigh wave at small $q$ and, in addition, a low-frequency optical mode whose frequency, $\omega_{\text{opt}}$ is proportional to $\sqrt{v}$ across the SBZ and that vanishes into the continuum at a critical wavenumber $q_0$.  The top surface, on the other hand, below the lowest bulk band only has a Rayleigh wave with the same velocity as that of the bottom surface, that disappears into the bulk continuum at a wavenumber that vanishes as $v\rightarrow 0$. The two zero-frequency modes of the topological $v=0$ lattice on the bottom surface are then the limits of the acoustic mode and the low-frequency optical mode.  At $v=0$, the bottom  of the band of bulk states, $\omega_{\text{band}}$, is proportional to  $q^2$ rather than $q$ as can be calculated from the envelope of the bulk dispersion, which has the form $\omega^2_{\text{bulk}} = (q_y^2-\beta q^2) +O(q^4)$ \cite{Kane2014,Lubensky2015} at small $q$.  Thus equating $\omega_{\text{opt}}$ to $\omega_{\text{band}}$ yields $q_0 \propto v^{1/4}$  in agreement with our numerical calculations. 

The finite $v$ frequency dispersions of surface states in both the $X_{\text{nt}}$ and $X_{\text{t}}$ lattices (both with p1 symmetry) depicted in Fig.~\ref{fig:bands} are in general different on the top and bottom surfaces, as one would expect because opposite surfaces in lattices with such low symmetry are not equivalent.  However, consistent with elastic theory, the small $q$ Rayleigh waves on both surfaces are the same and do not reflect p1 symmetry.  All of the higher frequency modes do however.  The high-$q$ frequencies of the acoustic modes of both lattices are different on  the  two surfaces as are all the higher-frequency optical surface modes [see SM].  

The approach of the finite-frequency phonons to the topological phonons is also reflected in their inverse penetration depths $\kappa$ shown in Figs.~\ref{fig:bands} (e) and (f). For both $X_{\text{nt}}$ and $X_{\text{t}}$, $\kappa$ is the same for $v>0$ at sufficiently small $q$ on the bottom and top surfaces as predicted by elastic theory, but differences between the bottom and top surfaces arise as $q$ becomes larger. As observed in the dispersion curves, the $v\to 0$ limit unfolds differently in the two lattices. For $X_{\text{nt}}$, the $\re(\kappa(q))$ curves of the two surfaces approach one another as $v$ vanishes and eventually become identical across the entire SBZ. 
For $X_{\text{t}}$, the $\re(\kappa(q))$ curves of the top surface terminate at values of $q$ that decrease with $v$ whereas the $\re(\kappa(q))$ curves of the acoustic and the lowest optical mode on the bottom surface approach each other  to produce a two-fold degenerate zero-frequency mode at $v=0$. Note that the penetration depth of the most dominant contribution to this mode diverges for $q \to 0$. The inset to Fig.~\ref{fig:bands} emphasizes the extremely small (but which we have verified is nonetheless positive) value of $\re(\kappa(q))$ throughout the region that the surface acoustic mode exists on the top surface indicating a very large penetration depth.

Our results for the GKL with bending forces are very similar [see SM]. The only notable difference is that the interaction strength $v$ is effectively larger than in the NNN model due to  factors mandated by the rotational invariance of the bending energies. Apart from that, the approach of the finite-frequency phonons to the topological phonons is qualitatively the same.

We now outline how these results were obtained.  The GKLs are derived from the standard kagome lattice (KL) by displacing \cite{Kane2014}  the 3 KL unit cell sites $\rv_1 = (0, 0)$, $\rv_2 = (1/2, 0)$, and $\rv_3 = (1/4, \sqrt{3}/4)$ by
\begin{subequations}
	\label{deformationDef}
	\begin{align}
	\delta \brm{r}_1 (X) & = \chi_1 \sqrt{3} \, \ev_1 - \chi_2\av_3\, ,
	\\
	\delta \brm{r}_2 (X) & = \chi_2 \sqrt{3} \, \ev_2 - \chi_3 \av_1\,  ,
	\\
	\delta \brm{r}_3 (X) & = \chi_3 \sqrt{3}\,  \ev_3 - \chi_1 \av_2\,,
	\end{align}
\end{subequations}
where $X\equiv (\chi_1, \chi_2, \chi_3)$ \footnote{Our  convention is equivalent to that of Ref.~[1,2] with a change in the signs of the $\chi$'s, so that the  topological polarization is $\Rv_T = -\frac{1}{2} \sum_\mu \Tv_\mu \, \text{sign} \chi_\mu$, where $\Tv_\mu$ are the primitive translation vectors.}. $\av_b$ are the normalized NN bond vectors of the KL: $\av_1 = (1, 0)$, $\av_2 = 1/2\, (-1, \sqrt{3})$,
$\av_3 = 1/2\,  (-1, -\sqrt{3})$. $\ev_b$ are unit vectors  perpendicular to the $\av_b$: $\ev_1 = (0, -1)$, $\ev_2 =1/2\,  (\sqrt{3}, 1)$, $\ev_3 = 1/2\, (-\sqrt{3}, 1)$. The displacements are designed \cite{Kane2014,Lubensky2015} so that making one of the $\chi_b$'s nonzero causes filaments (i.e., sample traversing straight lines of bonds) parallel to $\av_b$ to zigzag while keeping the remaining filaments straight. The crystallographic symmetry of the resulting GKL depends on $X$. For example, the twisted KL with $X = (\chi, \chi, \chi)$ (where $\chi$ is some reasonable positive or negative number) has p31m symmetry [see SM]. For $X = (0 ,\chi, \chi)$ and $X = (-\chi, \chi, \chi)$, the symmetry is reduced to cm and pm, respectively. Our generic GKLs  have deformation parameters $X = X_{\text{nt}} = (0.1, 0.15, 0.2)$ and $X = X_{\text{t}} = (0.1, 0.15, -0.2)$. We have chosen these parameters so that the resulting GKLs have the lowest possible (p1) symmetry, and  moderate distortions relative to the KL. Otherwise, these choices are arbitrary, and manifolds of alternative choices lead to qualitatively the same results. 

The vibrational modes of an elastic network are  governed by its dynamical matrix $\Dv$. In the bulk GKL, the equation of motion is simply 
$
\omega^2 \, \uv (\qv) = \Dv (\qv) \, \uv (\qv)  
$,
where $\uv = (u_{1x}, u_{1y}, u_{2x}, u_{2y}, u_{3x}, u_{3y})$ is the displacement vector of the basis sites, $\qv$ is the wave vector, and $\omega$ is the angular frequency. $\Dv = \Qv \Sv \Cv$ is the $6\times 6$ lattice dynamical matrix (for unit mass at sites), with $\Qv$ the equilibrium matrix,  $\Cv = \Qv^\dag$ the compatibility matrix, and $\Sv = \text{Diag} (1, 1, 1, v, v, v)$ the spring constant matrix (see Ref.~\cite{Lubensky2015} for background information). In the elastic (continuum) limit, $\uv$ turns into a 2-component displacement field and $\Dv$ turns into a  $2\times 2$ effective dynamical matrix.  Details about the dynamical matrices in the two theories are given in the SM.

To get a comprehensive picture, we use both lattice and elastic theory.  In our elastic theory, we adapt the standard textbook calculation~\cite{Landau-elasticity} of the decay lengths and sound velocities of acoustic surface phonons in isotropic continua to our anisotropic GLKs [see SM]. This approach applies only to the longest wavelength acoustic phonons. Our lattice-based calculations are a generalization to discrete lattices of the standard Rayleigh-wave continuum calculations \cite{Landau-elasticity}.  Like the latter calculations, they are done on semi-infinite systems that clearly separate top and bottom surfaces, yet they allow access to wave vectors ranging across the entire SBZ. To carry out our calculations, we break the lattice into one-cell-thick layers $L$, with $L=0$ the surface layer, $L=1$ the next layer into the bulk, and so on, stacked in the $y$-direction and with periodic boundary conditions along $x$. The equilibrium matrix has non-vanishing components $\Qv_{L,L} \equiv \Qv_{00}$ and $\Qv_{L,L-1} \equiv \Qv_{10}$ connecting sites in layer $L$ to bonds in layers $L$ and $L-1$, respectively; and the compatibility matrix has non-vanishing components $\Cv_{LL} \equiv \Cv_{00}$ and $\Cv_{L,L+1} \equiv \Cv_{01}$ connecting bonds in layer $L$ to sites in layers $L$ and $L+1$, respectively. The dynamical matrix then has components $\Dv_{L,L-1}=\Dv_{10}= \Qv_{10}\Sv\Cv_{00}$, $\Dv_{LL}=D_{00}=\Qv_{00}\Sv\Cv_{00}+\Qv_{10}\Sv \Cv_{01}$, and $\Dv_{L,L+1}=\Qv_{00}\Sv\Cv_{01}$.
The equation of motion for any layer $L>0$ then reads
\begin{equation}
\omega^2 \, \uv^L = \Dv_{10} \uv^{L-1} + \Dv_{00}\uv^{L} + \Dv_{01} \uv^{L+1} \, ,
\end{equation}
which is solved by $\uv^{L+1} = Z\, \uv^{L}$ provided that
\begin{equation}
\text{Det} \big[ \Dv_{10} Z^{-1} + \Dv_{00} + \Dv_{01} Z   
- \omega^2\,  \brm{\delta }\big] = 0 \, ,
\label{eq:Det-surface}
\end{equation}
where $\brm{\delta}$ is the unit matrix, and $Z$ determines the the inverse decay length  $\kappa$ in the $y$-direction via $Z = \exp (-\kappa)$ (with $\kappa$ in general complex). Solutions $Z(v, \omega, q)$ of Eq.~(\ref{eq:Det-surface}) come in pairs with reciprocal magnitude. Solutions with $|Z(v, \omega, q)| = 1$ correspond to bulk modes, whereas solutions with $|Z(v, \omega, q)| <1 (>1)$ decay away from the bottom (top) surface and correspond to surface modes. The points in $\omega$-$q$-space where bulk modes exist, i.e., points for which there is at least one pair of solutions with magnitude 1, form bands akin to the projected band structures in electronic systems (see Fig.~\ref{fig:bands}). Surface modes can exist only within the bulk band gaps as the solutions of the equations of motion must obey the conditions imposed by the surface. Sites 1 and 2 of the surface unit cell lie directly in the free surface (note that site 3 does not). The force on these surface sites comes only from NN and NNN bonds $1$ to $4$ [Fig.~\ref{fig:genKagomeNNN}] in the zeroth layer, and as a result for the free boundary condition we impose, the first four components of the force vector satisfy $\fv^0  =  \Dv_{00} \uv^{0} +  \Dv_{01} \uv^{1}  = \omega^2 \uv^0$. For $v>0$, there are a total of eight zeros at any point in $\omega$-$q$-space. This implies that, at any point in a band gap, there are four modes with $|Z(v, \omega, q)| <  1 (>1)$ that decay away from the bottom (top) surface. The boundary conditions can be satisfied by superimposing these decaying modes, 
\begin{align}
\label{genForm}
\uv^L = \sum_{n=1}^4 A_n \wv_n Z_n^L e^{i q x-i\omega t} \, ,
\end{align}
where the $A_n$ are mode amplitudes and  $\wv_n = \wv_n  (v, \omega, q, Z_n)$ are polarization vectors. The band gap points for which the determinant of the $4\times 4$ boundary matrix $\Bv$, defined by
\begin{align}
\label{boundaryMat}
B_{mn} =   \sum_{k=1}^6 \left[\Dv_{00} +  \Dv_{01} Z_n  - \omega^2 \brm{\delta} \right]_{mk} w_{n, k} \, ,
\end{align}
vanishes determine the dispersion relation of the surface modes. To find these points, we use the standard secant-method for computing zeros with an array of starting points that sweeps the band gaps.

Modern $3d$ printing and cutting techniques now produce bespoke materials, including regular lattices, with almost arbitrary designs.  In particular, these techniques can produce mechanical lattices, whose geometry is almost identical to isostatic mechanical NN topological lattices. To fully understand and control these lattices, it is important to know how their properties - elastic energy, bulk- and surface- mode structure, etc. - differ from those of the ideal NN isostatic lattice. Our formalism treats semi-infinite systems exactly and can easily be used to calculate linearized response, for example to a localized force at a surface. 

The main result of the present work is the unravelling of the apparent dilemma of topological lattices where topological-phonon theory predicts for the example we are studying two soft surface modes on one surface (soft, bottom) and zero on the other (hard, top) whereas elasticity theory mandates that there be one Rayleigh wave per surface wavenumber on each surface and that the two waves have equal speeds.Ê Our work shows that the resolution of the dilemma is as follows: as $v \to 0$, the domain of existence of the Rayleigh wave on the top surface shrinks to zero. On the bottom surface, there is a low-energy optical surface mode, whose domain grows to the full SBZ and which approaches the bottom surface Rayleigh wave as $v \to 0$. These two together produce the two surface-zero modes predicted by topological-phonon theory.   

Our results provide guidance for interpreting results of experiments on metamaterials targeting topological phonons. Reference \cite{MaZhou2018} reports experiments and finite element analysis on kagome-like lattices that show an asymmetric bulk phonon spectrum in a topological lattice but a symmetric one in a non-topological lattice. They verify the existence, in the same geometry we study, of the two low-energy surface modes on the soft surface that emerge from the two zero modes of the ideal topological lattice which they interpret as ``an interesting departure from the conventional case of Rayleigh waves''. Curiously neither the finite element analysis nor the measurements show any evidence of the acoustic Rayleigh wave on the hard surface mandated by elasticity theory. It would be interesting to see additional experiments that specifically target the evolution of the hard surface Rayleigh wave with increasing bending rigidity.

\begin{acknowledgments}	
This work was supported by the NSF under No.~DMR-1104701 (OS,
TCL), No.~DMR-1120901 (OS, TCL), and No.~DMR-1720530 (OS, TCL). TCL was supported  by a Simons Fellows grant.
\end{acknowledgments}

\newpage

\newpage
\section{Supplemental Material}

\subsection{Symmetries of the GKL}
\begin{figure*}[ptb]
	\includegraphics[]{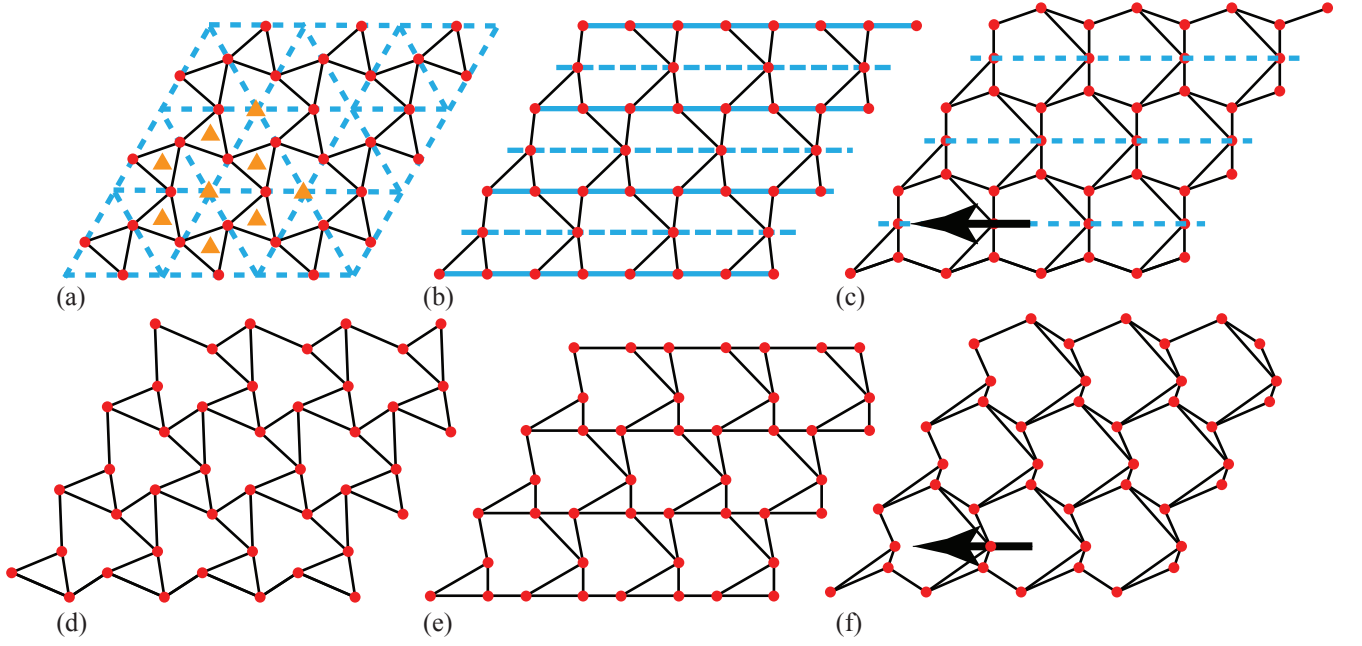}
		\caption{GKLs with different symmetries: (a) Non-topological lattice with $X=(-0.1,-0.1,-0.1)$ with p31m symmetry - wallpaper (WP) group 14; the dashed blue lines indicate mirror lines and the orange triangles three-fold rotation axes. (b) Transition  lattice with $X=(0,-0.1,-0.1)$ with cm symmetry - WP group 5; it has both mirror lines  (dashed  blue lines) and reflection glide lines (full blue line). (c) Topological lattice with $X=(0.1,-0.1,-0.1)$ with pm symmetry - WP group 3; it has mirror lines but no glide lines. (d) $X=(-0.15,-0.1,-0.2)$. (e) $X=(0,-0.1,-0.2)$. (f) $X=(0.15,-0.1,-0.2)$.  (d) to (f) all have the lowest p1 symmetry, even though (d) is ``non-topological", (e) is a critical lattice, and (f) is topological.}%
	\label{fig:lattice-symmetries}%
\end{figure*}

The topological properties of Maxwell lattices, and our GKLs in particular, are not determined by their geometric  symmetry, even though there are symmetry changes for lattices with $X=(\chi_1, \chi,\chi)$ as $\chi_1$ changes sign as can be seen from (a) to (c) in Fig.~\ref{fig:lattice-symmetries}. These lattices, the gapped non-topological [$X=(-0.1,-0.1,-0.1)$] and topological [$X=(0.1,-0.1,-0.1)$] lattices and the critical  [$X=(0,-0.1,=0.1$)] lattice in which the gaps along $q_y$ vanish, all have different symmetries.  All three of these lattices can, however, be  continuously distorted into ``generic" lattices with the lowest polar p1 symmetry, as shown in Figs.~\ref{fig:lattice-symmetries} (d) to (f), without changing their gap structure simply by allowing the magnitudes of $\chi_1$, $\chi_2$, and $\chi_3$ to be different.  It should be noted that all topological  lattices with a non-vanishing topological polarization have a geometric polar symmetry [wallpaper groups p1 or pm] but both non-topological and critical lattices can  also have this symmetry.  In the main  text, we focused on the surface band structure of generic lattices.

\subsection{GKL with NNN stretching forces}

\subsubsection{Model energy}
To adapt the GKLs to the superisostatic situation typically found in the lab, we augment them here with NNN springs. This leads to the ball-and-spring model elastic energy
\begin{align}
E = \frac{1}{2} \sum_{NN} \left( s_b^{NN} \right)^2 +
\frac{v}{2} \sum_{NNN} \left( s_b^{NNN} \right)^2 \, ,
\end{align}
where the first sum runs over the 6 NN bonds and the second sum over the 6 NNN bonds of the unit cell shown in Fig.~1 (a) of the main paper.
\begin{align}
s_b^{NN}  = \uv_b \cdot \av_b\, ,
\end{align}
is the stretch of NN bond $b$, where $\uv_b = \uv_i-\uv_j$ is the difference in the elastic displacements of lattice sites $i$ and $j$ connected by that bond which has a normalized bond vector $\av_b$. The NNN-bond stretch $s_b^{NNN}$ is defined in a similar, obvious manner. For simplicity, we have set the spring constant of the NN bonds and the masses of the sites equal to 1. 

\subsubsection{Lattice theory -- equilibrium, compatibility and dynamical matrixes}
The equilibrium, compatibility and dynamical matrixes are elementary to the lattice description of elastic networks. For any $d$-dimensional central-force elastic network with $n$ sites and $n_B$ bonds, the $n_B \times dn$ compatibility matrix $\brm{C}(\qv)$
relates bond displacements $\uv(\qv)$ to bond extensions
$\ev(\qv)$ via $\brm{C}(\qv) \uv(\qv) = \ev(\qv)$. The null space of $\Cv(\qv)$ constitutes the zero modes of the network. The
$dn\times n_B$ equilibrium matrix $\brm{Q}(\qv)=\Cv^\dag(\qv)$
relates bond tensions $\tv(\qv)$ to site forces $\fv(\qv)$ via
$\Qv(\qv) \tv(\qv) = \fv(\qv)$. Its null space constitutes the states of self-stress of the network.  The $dn\times dn$ dynamical matrix governing the phonon spectrum is related to the equilibrium and  compatibility matrixes by $\Dv(\qv) =\Qv(\qv)\Sv \Cv(\qv)$, where $\Sv$ is the spring constant matrix.

The bulk compatibility matrix of our model lattice with NNN bonds
reads
\begin{widetext}
\begin{align}
\brm{C} (\brm{q}) =
 \left(
\begin{array}{cccccc}
a_{1,x} & a_{1,y}  & -a_{1,x} & -a_{1,x} & 0 & 0
\\
0 & 0 & a_{2,x} & a_{2,y} & -a_{2,x} & -a_{2,y}
\\
-a_{3,x} & -a_{3,y}  & 0 & 0 & a_{3,x} & a_{3,y}
\\
-a_{4,x} e^{-i \qv \cdot \Tv_1}& -a_{4,y} e^{-i \qv \cdot \Tv_1}  & a_{4,x} & a_{4,y} & 0 & 0
\\
0 & 0 & -a_{5,x} e^{i \qv \cdot \Tv_3} & -a_{5,x} e^{i \qv \cdot \Tv_3} & a_{5,x} & a_{5,y}  
\\
-a_{6,x} e^{-i \qv \cdot \Tv_2}  & -a_{6,y} e^{-i \qv \cdot \Tv_2} & 0 & 0 & a_{6,x} & a_{6,y} 
\\
b_{1,x} & b_{1,y} &  - b_{1,x} e^{i \qv \cdot \Tv_3} & - b_{1,y} e^{i \qv \cdot \Tv_3} & 0 & 0
\\
0 & 0 & b_{2,x} & b_{2,y} &  - b_{2,x} e^{-i \qv \cdot \Tv_1} & - b_{2,y} e^{-i \qv \cdot \Tv_1}
\\
b_{3,x} & b_{3,y} & 0 & 0 & - b_{3,x} e^{i \qv \cdot \Tv_1} & - b_{3,y} e^{i \qv \cdot \Tv_1}
\\
- b_{4,x} e^{-i \qv \cdot \Tv_2} & - b_{4,y} e^{-i \qv \cdot \Tv_2} & b_{4,x} & b_{4,y} & 0 & 0
\\
0 & 0 & - b_{5,x} e^{-i \qv \cdot \Tv_2} & - b_{5,y} e^{-i \qv \cdot \Tv_2} & b_{5,x} & b_{5,y}
\\
- b_{6,x} e^{i \qv \cdot \Tv_3} & - b_{6,y} e^{i \qv \cdot \Tv_3} & 0 & 0 & b_{6,x} & b_{6,y}
\end{array}
\right) 
\end{align}
\end{widetext}
in $\brm{q}$-space,  where $\brm{T}_1 = (1,0)$ and $\brm{T}_2 =  \frac{1}{2}(1, \sqrt{3})$ are the primitive translation vectors we are using, and $\brm{T}_3 = \brm{T}_1- \brm{T}_2= \frac{1}{2}(1, -\sqrt{3})$. $\brm{T}_1$, $\brm{T}_2$, and $\brm{T}_3$ are chosen so that their sum is zero. Note that the primitive translation vectors are independent of $X$. $\brm{a}_b$ and $\brm{b}_b$ are the normalized NN and NNN bond vectors of the GKL, respectively. These depend on the deformation parameters  $X$. For $X= X_{\text{nt}}$, for example, 
\begin{subequations}
\begin{align}
\brm{a}_1 &=  \frac{1}{\sqrt{31}} \left(3 \sqrt3, 2 \right) , \quad  \brm{a}_2 = \frac{1}{2 \sqrt{19}} \left(-7, 3\sqrt{3} \right)  ,
\\
\brm{a}_3 &=  \frac{1}{2 \sqrt{43}} \left(\sqrt{3}, -13  \right), \quad  \brm{a}_4 = \frac{1}{\sqrt{133}} \left(11,-2 \sqrt{3}\right)  ,
\\
\brm{a}_5 &=  \frac{1}{2 \sqrt{91}} \left(1, 11 \sqrt{3}  \right), \quad  \brm{a}_6 = \frac{1}{26} \left(23, 7 \sqrt{3}\right)  ,
\end{align}
\end{subequations}
and
\begin{subequations}
\begin{align}
\brm{b}_1 &=  \frac{1}{\sqrt{433}} \left(-1 , 12 \sqrt3 \right) , \quad  \brm{b}_2 = \frac{1}{2\sqrt{151}} \left( 19, 9 \sqrt{3} \right)  ,
\\
\brm{b}_3 &=  \frac{1}{2\sqrt{589}} \left(- 43, 13 \sqrt3  \right), \quad  \brm{b}_4 = \frac{1}{\sqrt{193}} \left(1,8 \sqrt{3}\right)  ,
\\
\brm{b}_5 &=  \frac{1}{2 \sqrt{511}} \left(41, 11 \sqrt{3}  \right), \quad  \brm{b}_6 = \frac{1}{2 \sqrt{109}} \left(-17, 7 \sqrt{3}\right)  .
\end{align}
\end{subequations}
Calculating the nullspaces of the equilibrium and compatibility matrixes for $X_{\text{nt}}$ and $X_{\text{t}}$, we find that there are 8 states of self-stress and the 2 inevitable trivial zero modes for $\brm{q} =\brm{0}$ which is consistent with the Maxwell counting. The dynamical matrix of the lattice theory is readily obtained by taking the product of the equilibrium and compatibility matrixes.

In the presence of a planar surface, it is useful to decompose equilibrium, compatibility and dynamical matrixes into layer matrixes describing springs respectively connecting sites in the same and ones in different surface-parallel layers. For our choice of having a free surface parallel to the $x$-direction, we have
\begin{widetext}
\begin{align}
\brm{C}_{00} (q) = 
 \left(
\begin{array}{cccccc}
a_{1,x} & a_{1,y}  & -a_{1,x} & -a_{1,x} & 0 & 0
\\
0 & 0 & a_{2,x} & a_{2,y} & -a_{2,x} & -a_{2,y}
\\
-a_{3,x} & -a_{3,y}  & 0 & 0 & a_{3,x} & a_{3,y}
\\
-a_{4,x} e^{-i q}& -a_{4,y} e^{-i q}  & a_{4,x} & a_{4,y} & 0 & 0
\\
0 & 0 & 0 & 0 & a_{5,x} & a_{5,y}  
\\
0  & 0 & 0 & 0 & a_{6,x} & a_{6,y} 
\\
b_{1,x} & b_{1,y} &  0 & 0 & 0 & 0
\\
0 & 0 & b_{2,x} & b_{2,y} &  - b_{2,x} e^{-i q} & - b_{2,y} e^{-i q}
\\
b_{3,x} & b_{3,y} & 0 & 0 & - b_{3,x} e^{i q} & - b_{3,y} e^{i q}
\\
0 & 0 & b_{4,x} & b_{4,y} & 0 & 0
\\
0 & 0 & 0 & 0 & b_{5,x} & b_{5,y}
\\
0& 0 & 0 & 0 & b_{6,x} & b_{6,y}
\end{array}
\right) 
\end{align}
for the intra-layer compatibility matrix and
\begin{align}
\brm{C}_{01} (\brm{q}) =
 \left(
\begin{array}{cccccc}
0& 0  & 0 & 0 & 0 & 0
\\
0 & 0 & 0 & 0 & 0 &0
\\
0 & 0  & 0 & 0 &0 & 0
\\
0& 0  & 0 & 0 & 0 & 0
\\
0 & 0 & -a_{5,x} e^{i q/2} & -a_{5,x} e^{i q/2} & 0 & 0 
\\
-a_{6,x} e^{-i q/2}  & -a_{6,y} e^{-i q/2} & 0 & 0 &0 & 0 
\\
0 & 0 &  - b_{1,x} e^{i q/2} & - b_{1,y} e^{i q/2} & 0 & 0
\\
0 & 0 & 0 & 0 &  0 & 0
\\
0 & 0 & 0 & 0 & 0 & 0
\\
- b_{4,x} e^{-i q/2} & - b_{4,y} e^{-i q/2} & 0 & 0 & 0 & 0
\\
0 & 0 & - b_{5,x} e^{-i q/2} & - b_{5,y} e^{-i q/2} & 0 & 0
\\
- b_{6,x} e^{i q/2} & - b_{6,y} e^{i q/2} & 0 & 0 & 0 & 0
\end{array}
\right) 
\end{align}
for the extra-layer compatibility matrix, where $q=q_x$.
\end{widetext}

\subsubsection{Elastic theory -- Lagrange elastic energy}
Under imposed external strain, basis sites $\alpha$ undergo
displacements $u_{\alpha,i}= \eta_{ij} x_{\alpha,i} + \delta
u_{\alpha,i}$ for $i=x,y$, where $\eta_{ij}$ is the imposed
macroscpic deformation, and $\delta u_{\alpha,i}$ is the nonaffine part of the displacement. Minimizing our model elastic energy over $\delta u_{\alpha,i}$, we obtain an effective elastic energy density that can be expressed in terms as the usual Lagrange strain tensor $u_{ij}$. 

For the conformations of the GKL with higher symmetry, the effective elastic energy can be very simple. The GKL with $X= (\chi, \chi, \chi)$ , for example, corresponds to the twisted kagome lattice which is macroscopically isotropic. Hence is Lagrange energy density is of the form
\begin{align}
\label{genElastEnIso}
f = \frac{\lambda}{2}  \, u_{ii}^2 + \mu \, u_{ij}u_{ij} \, .
\end{align}
The Lame coefficients of this lattice with $\chi = 0.1$ are, e.g.,  given by
\begin{subequations}
\begin{align}
\label{lameX_2}
\lambda &= \frac{3 (-3 +31v+84v^2)}{16(3 + 28 v)} \, ,
\\
\mu &= \frac{3 (3 +37v+84v^2)}{16(3 + 28 v)} \, .
\end{align}
\end{subequations}
Note that the bulk modulus $B = \lambda + \mu$ vanishes for $v\to 0$ as it should for the twisted kagome lattice without NNN bonds.

For our generic lattices $X_{\text{nt}}$ and $X_{\text{t}}$, the Lagrange energy density is considerably more complicated because there are six independent elastic constants:
\begin{align}
\label{genElastEnXMinus3}
f &= \frac{1}{2}  K_{11} \, u_{xx}^2 +\frac{1}{2}  K_{22} \, u_{yy}^2 + K_{12} \,u_{xx}u_{yy}+ 2 K_{33}\, u_{xy}^2 +
\nonumber \\
&  + 2K_{13} \, u_{xx} u_{xy}+ 2K_{23} \, u_{yy} u_{xy}\, .
\end{align}
After Fourier transformation of $f$, we can straightforwardly extract the dynamical matrix for $X_{\text{nt}}$ and $X_{\text{t}}$ in elastic theory by  taking second derivatives with respect to the components of the elastic displacement.

For $X_{\text{nt}}$, the six elastic constants are given by
\begin{widetext}
\begin{subequations}
\begin{align}
\label{genElastEnX1GenElastConst}
B \, K_{11} & = \textstyle{\frac{1}{4}} \Big(64066387072758047378597 + 826332597356205448762093 v + 
   2787778202630610433014742 v^2 
   \nonumber\\
   &+ 3381681827843928199167638 v^3 + 
   1357752557780034687526437 v^4 + 12534744296675727990045 v^5\Big)\, ,    
\\
B \, K_{22}&= \textstyle{\frac{1}{4}} \Big(-31741479400541232633129 + 86184269379066175180667 v + 
   740891956058650166834770 v^2 
    \nonumber\\
   &+ 1115965371795836672225418 v^3 + 
   493549308640875173434839 v^4 + 5898796907214956230395 v^5\Big) \, ,
   \\   
B \, K_{33}&= - \textstyle{\frac{\sqrt{3}}{4}} \Big(583428271675582112339 - 4157773102727939151705 v - 
   5838777757209331264674 v^2 
    \nonumber\\
   &+ 5156348205989863308398 v^3 + 
   7211220231643636602303 v^4 + 1105991231576795293275 v^5\Big) \, , 
\\
B \, K_{12} &= \textstyle{\frac{3}{4}} \Big(6010787258233919572719 + 250265267692922730445327 v + 
   960047508094656245737010 v^2 
    \nonumber\\
   &+ 1200964451099927268925858 v^3 + 
   485931485533103278158399 v^4 + 4410736097612180096943 v^5 \big) \, ,
\\
B \, K_{13}&= \textstyle{\frac{\sqrt{3}}{4}}  \Big(-4323244148862284803185 - 13716229149805239387961 v - 
   13548108764853625951234 v^2 
    \nonumber\\
   &- 2468306508061388354586 v^3 + 
   2760356898057381688611 v^4 + 1222235230936930676739 v^5\Big) \, ,
\\
B \, K_{23} &=\textstyle{\frac{1}{2}}\Big(749708200455156127 + 290104040614985122663771 v + 
 983776712630814114256758 v^2 
  \nonumber\\
   &+ 1213564070972292895811718 v^3 + 
 488017109696386616047611 v^4 + 1164113231256862985007 v^5 \Big) \, ,
\end{align}
with $B$ being an abbreviation for
\begin{align}
\label{Bdef}
B  &=   \textstyle{\frac{8}{3}}\Big(16664365670864352850787 + 110800511173844068807831 v + 
   174745551136457039371671 v^2 
    \nonumber\\
   &+ 80387187958650600842997 v^3 + 
   796574499957021428370 v^4\Big) \, .
\end{align}
\end{subequations}
For $X_{\text{t}}$, the elastic constants read
\begin{subequations}
\begin{align}
\label{genElastEnX-3GenElastConst}
B \, K_{11} & =  \textstyle{\frac{1}{12}}\Big(22968827725654518158087 + 608079378075643941524223 v + 
  2315015433759049440961746 v^2 
  \nonumber\\
   &+ 2973129913162249070121170 v^3 + 
  1310973960535841431703559 v^4 + 82311191396248539177999 v^5\Big)\, , 
\\
B \, K_{22} &=  \textstyle{\frac{1}{4}}\Big(1871876344007820777263 + 14543158219536761721619 v + 
  81741217587888066658722 v^2 
  \nonumber\\
   &+ 151989412941048260843482 v^3 + 
  103607540769847961350959 v^4 + 20808603404027818221843 v^5\Big) \, ,
   \\
B \, K_{33}&= - \textstyle{\frac{7}{4 \sqrt{3}}}\Big(-392348470155895440889 + 10040431773637819064331 v + 
  27035407302484224017574 v^2 
  \nonumber\\
   &+ 23376116240867084249846 v^3 + 
  9184980163653051188739 v^4 + 1988538118416508451247 v^5\Big)\, , 
\\
B \,K_{12}&= \textstyle{\frac{1}{4}}\Big(4874229176204078380053 + 120132238311877217826325 v + 
  491483797429715034773510 v^2 
  \nonumber\\
   &+ 663232825986241130284310 v^3 + 
  302514287746820382533125 v^4 + 20361336925819867576053 v^5\Big)\, ,
\\
B \, K_{13} &= -  \textstyle{\frac{7 \sqrt{3}}{4}}\Big(28114485172684996111 + 164672570321857924487 v + 
  3998248601896106470974 v^2 
  \nonumber\\
   &+ 7628970939921065737510 v^3 + 
  3671022783308271259491 v^4 + 44726647820795064579 v^5\Big) \, ,
\\
B \, K_{23} &= \textstyle{\frac{1}{2}} \Big( 448577440501265433111 + 93813264622554016682451 v + 
 294569403936461829806886 v^2 
 \nonumber\\
  &+ 313555729213981632060662 v^3 + 121316827653978765464915 v^4 + 9086355057275203961799 v^5\,\Big) , 
\end{align}with $B$ abbreviating
\begin{align}
\label{BdefXt}
B  &=   \textstyle{\frac{8}{3}}\Big(6601908583705608497051 + 32777957696830528117583 v + 
  48530927362979586968143 v^2 
  \nonumber\\
   &+ 24385808037090842131741 v^3 + 
  2341843958115406784130 v^4\Big) \, .
\end{align}
\end{subequations}
\end{widetext}

\subsubsection{Calculation of elastic Rayleigh waves in general anisotropic crystals}

We seek Rayleigh waves on edges parallel to the $x$-axis and decaying exponentially into the bulk for $y>0$.  The elastic dynamical matrix, $\overleftrightarrow{D}$, is homogeneous in $q_x$ and $q_y$, and we can assume that $q_y = p q_x \equiv p q$, where $p$ must have a positive imaginary part. In this case, we can scale $\overleftrightarrow{D}$ via $\overleftrightarrow{D}=K_{11} q^2 \overleftrightarrow{R}$, where the components of $ \overleftrightarrow{R}$ are
\begin{subequations}
\begin{align}
R_{11} & = 1+ 2 k_{13} p + 2 k_{33} p^2 \\
R_{22} & = k_{33} + 2 k_{23} p + k_{22} p^2  \\
R_{12} & = R_{21} = k_{13} + (k_{12}+ k_{22}) p + k_{23} p^2 ,
\end{align}
\end{subequations}
where $k_{ab} = K_{ab}/K_{11}$.  Then, 
\begin{equation}
\overleftrightarrow{R} \vec{u} = s \vec{u} ,
\label{eq:eigen}
\end{equation}
where $\vec{u} = (u_x, u_y)$, $s =\omega^2/(K_{11} q^2)$, and where
\begin{equation}
\det[\overleftrightarrow{R}- s \overleftrightarrow{I}]= s^2 -(R_{11} + R_{22}) s + \det[\overleftrightarrow{R}] = 0,
\end{equation}
determines the relation between $p$ and $\omega$.  This is a quartic equation in $p$ whose solutions are either real or part of a complex-conjugate pair.  Two decaying solutions, i.e., solutions with positive imaginary parts for $p$, are needed to meet the decay constraint and the surface boundary conditions, so in parameter regions where Rayleigh waves exist, there are two complex conjugate pairs.  This means that two solutions have positive imaginary parts and two have identical negative imaginary parts implying that the Rayleigh waves on opposite surfaces will have exactly the same energy and penetration depths in spite of the fact that opposite surfaces are not equivalent in systems with polar p1 symmetry.

To determine $s$, and thus $p$, we impose the boundary condition of zero stress at the edge $y=0$:
\begin{subequations}
\begin{align}
S_{xy} & =(k_{12} + p k_{23}) u_x + (k_{23} + pk_{22}) u_y = 0  \\
S_{yy} & = (k_{13} + p k_{33}) u_x + (k_{33} + p k_{23}) u_y = 0 ,
\label{eq:stress1}
\end{align}
\end{subequations}
where $S_{ab} = \sigma_{ab}/ (iq  K_{11})$ is the reduced stress tensor.  The solutions to Eq.~(\ref{eq:eigen}) are 
\begin{align}
u_a(q,\omega,y) & = A v_a^{(1)} (\omega,q) e^{i p_1(q,\omega) y} + B v_a^{(2)} (\omega,q) e^{i p_2(q,\omega) y}\, ,
\end{align} 
where $a=x,y$,
which when inserted into Eq.~(\ref{eq:stress1}) yield
\begin{equation}
\overleftrightarrow{M}\cdot 
\begin{pmatrix}
	A\\
	B
\end{pmatrix} 
=
\begin{pmatrix}
	0\\
	0
\end{pmatrix}  ,
\end{equation}
where
\begin{subequations}
\begin{align}
M_{11} & = (k_{12} + p_1 k_{23}) v_x^{(1)} + (k_{23} + p_1 k_{22})v_y^{(1)} \, ,  \\
M_{12} & = (k_{12}+p_2 k_{23}) v_x^{(2)} + (k_{23}+ p_2 k_{22}) v_y^{(2)} \, , \\
M_{21} & = (k_{13} + p_1 k_{33}) v_x^{(1)}+ (k_{33}+ p_1 k_{23}) v_y^{(1)} \, ,\\
M_{22} & = (k_{13}+p_2 k_{33}) v_x^{(2)} + (k_{33}+ p_2 k_{23}) v_y^{(2)} .
\end{align}
\end{subequations}
The Rayleigh wave sound speed is determined by $| \det \overleftrightarrow{M}(\omega, p_1(\omega), p_2(\omega),q)|=0$.  This program is easily implemented numerically.

\subsubsection{Top surface and deformation parameters with flipped signs}
\begin{figure}[ptb]
\includegraphics[width=5.8cm]{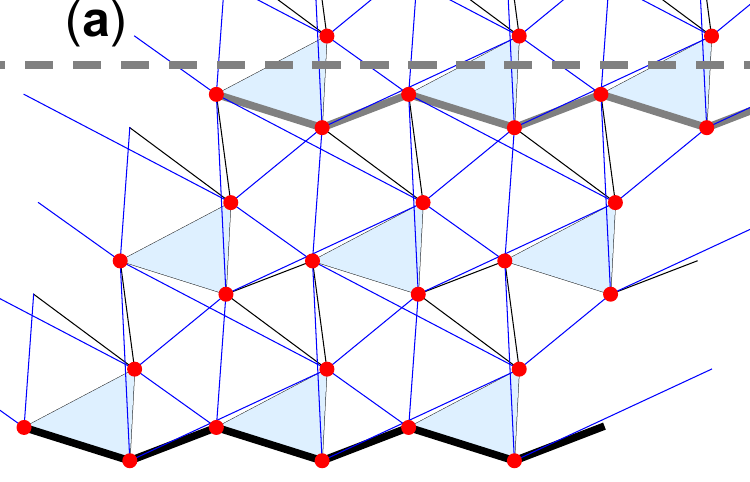}
\\
\vspace{0.5cm}
\includegraphics[width=5.8cm]{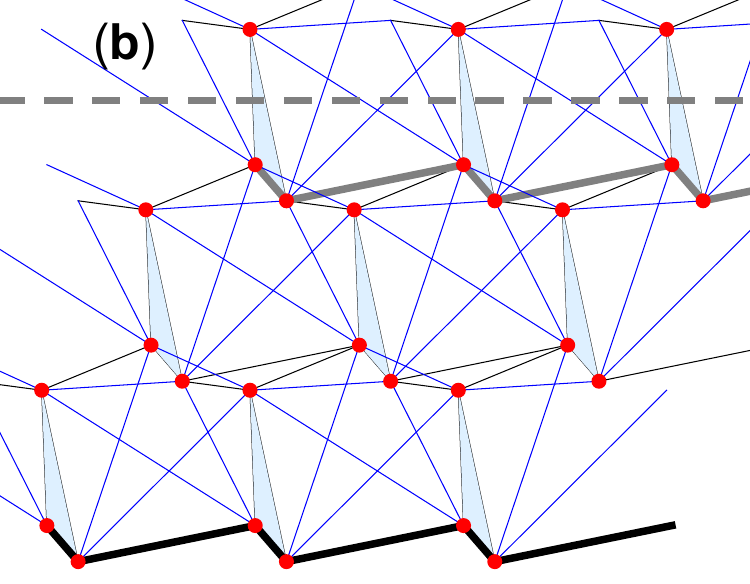}
\caption{$\bar{X}_{\text{nt}}$ (a) and $\bar{X}_{\text{t}}$ (b) conformation with a bottom surface parallel to the $x$-direction.}%
\label{fig:XBar}%
\end{figure}

The top surface of our model network can be conveniently studied by flipping the signs of the deformation parameters $X$ while keeping the surface at the bottom. For the top surface of $X_{\text{nt}}$, we can instead do our actual calculation for the bottom surface of $\bar{X}_{\text{nt}} = (-0.1, -0.15, -0.2)$. In the topological case, we can likewise use $\bar{X}_{\text{t}} = (-0.1, -0.15, 0.2)$ instead of $X_{\text{t}}$. Comparison of Fig.~\ref{fig:XBar} (a) [(b)] with Fig.~1 (c) [(d)] of our main paper demonstrates that the bottom surface of $\bar{X}_{\text{nt}}$ [$\bar{X}_{\text{t}}$] is equivalent to the top surface of $X_{\text{nt}}$ [$X_{\text{nt}}$] up to an inconsequential rotation of the entire system by $180^\circ$. 

\subsubsection{Full band structure}
\begin{figure}[ptb]
	\includegraphics[width=6.0cm]{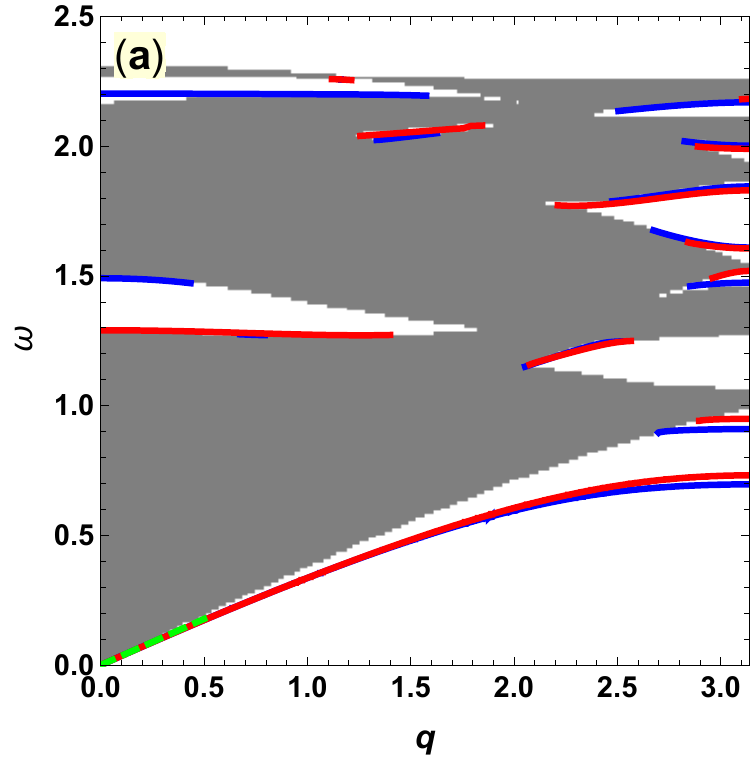}
	\includegraphics[width=6.0cm]{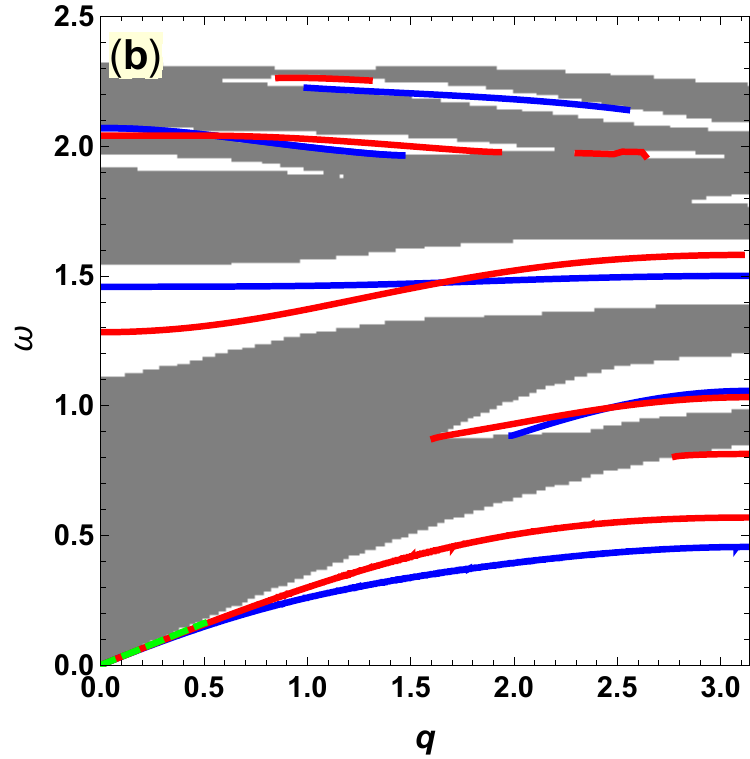}
\caption{The full surface band structure for (a) $X_{\text{nt}}$ and (b) $X_{\text{t}}$ at $v=0.5$.  Their color code is the same as that in Fig.~2 of the main paper.}
	\label{fig:bandsV05}%
\end{figure}

In the main text, our focus lies on the lowest frequency surface modes that approach that become topological zero modes in the isostatic limit. Our lattice calculation approach, however, allows us to go beyond to low-frequency limit and calculate the full surface mode structure. Figure~\ref{fig:bandsV05} presents as an example the full surface mode structure of the NNN GKL with $X_{\text{nt}}$ and $X_{\text{t}}$ at $v=0.5$. Note that the dispersions of the optical surface modes are different for the bottom and top surfaces.

\subsubsection{Inverse penetration depth}
\begin{figure}[ptb]
	\includegraphics[width=6.0cm]{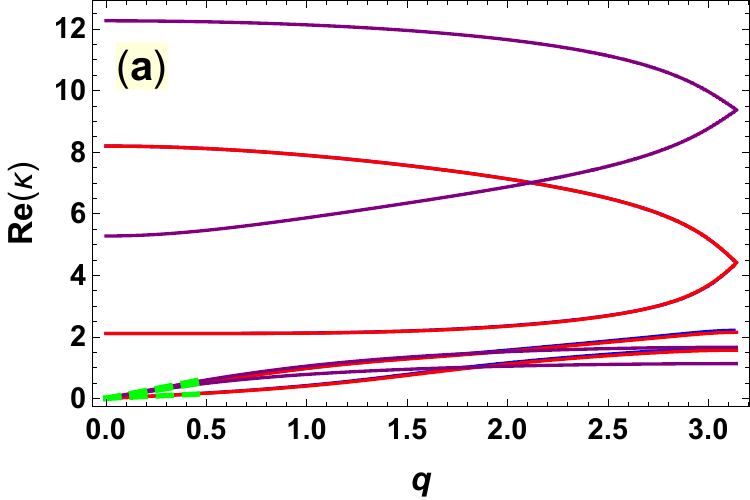}
	\includegraphics[width=6.0cm]{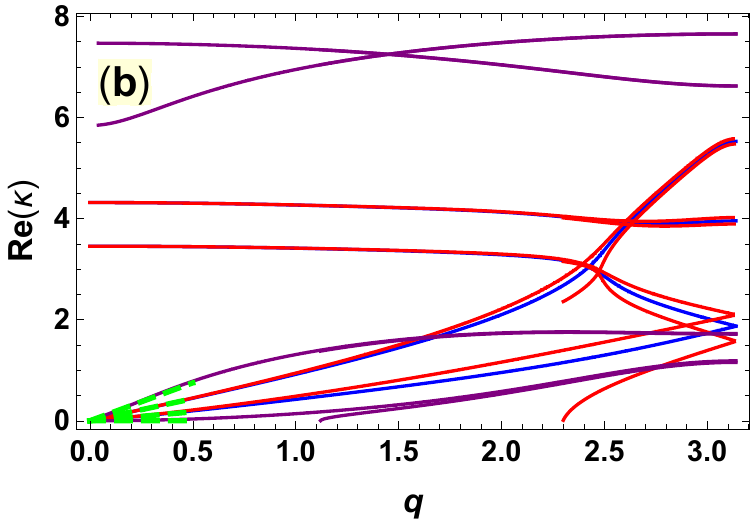}
\caption{The full set of $\re(\kappa(q))$ curves for (a) the acoustical surface modes of $X_{\text{nt}}$ and for (b) acoustical surface modes and lowest optical bottom surface mode of $X_{\text{t}}$.  The color code is the same as that in Fig.~2 (e) and (f) of the main paper.}
	\label{fig:fullReKappa}%
\end{figure}

Each of our surface modes consist of a superimposition of four normal modes that decay away from the respective surface and hence each of our surface modes is associated with four $\re(\kappa(q))$ curves. In Fig. 2 (e) and (f) of the main text, we focus on the two longest ranging contributions to each surface mode, i.e., we display only the two lowest $\re(\kappa(q))$ curves for each surface mode to make the plots less busy. For the sake of completeness, we show here in Fig.~\ref{fig:fullReKappa} the full set of $\re(\kappa(q))$ curves pertaining to the surface modes in Fig. 2 (a) to (d) of the main text.

\subsection{GKL with bending forces}

\subsubsection{Model energy}
In the usual GKL, the lattice sites act as free hinges, i.e., there is no preferred angle between any pair of bonds that meet at a given site. Here, we extend the GKL to include bending energies that penalize deviations of bond-pair angles from their equilibrium values, see Fig.~1(b) of our main text. Our model elastic energy reads
\begin{align}
\label{modelEn}
E = \frac{1}{2} \sum_{b=1}^6  s_b^2 +
\frac{v}{2} \sum_{B=1}^{12}  \theta_B^2 \, ,
\end{align}
where the NN stretching contribution with the bond stretch
$s_b = \uv_b \cdot \hat{\av}_b$,
where $\uv_b = \uv_i-\uv_j$ is the difference in the elastic displacements of lattice sites $i$ and $j$ connected by bond $b$, is  identical to that of our model with added NNN forces. 
In the second term, the bending contribution, the sum runs over the 12 bond pairs associated with the angles defined in Fig.~1 (b) of our main text. $\theta_B$ measures the deviation of the angle of bond pair $B$ from its equilibrium value $\theta_B^0$, and $v$ is the bending stiffness. The specific form of $\theta_B$ depends on the value of $\theta_B^0$. For $\theta_B^0 =0$, i.e., for a pair of bonds $b$ and  $b^\prime$ that is straight in equilibrium, 
\begin{align}
\theta_B = \wv_{b, \perp} -  \wv_{b^\prime, \perp}\, ,
\label{straightAngle}
\end{align}
with $\wv_{b, \perp} = P_b \wv_{b}$. Here, $\wv_{b} = \uv_b/|\xv_b|$, where $|\xv_b|$ is the equilibrium length of the bond and $P_b = \brm{\delta} - a_b a_b$, with $\brm{\delta}$ the unit matrix, the projector on the direction perpendicular to it. For the generic GKLs that we focus on in our present work, all bond pairs are bent to some degree at equilibrium, $\theta_B^0 >0$, so that
\begin{align}
\theta_B = \frac{ \wv_{b, \perp} \cdot a_{b^\prime} + \wv_{b^\prime, \perp} \cdot a_{b}}{\sin \theta_B^0} \, .
\label{crookedAngle}
\end{align} 
for all bond pairs. Note that, by construction,  $\theta_B$ is invariant under global rotations, and that the four  $\theta_B$'s about any given node sum up to zero.

For our actual calculations, it is more convenient to rewrite the model elastic energy as
\begin{align}
\label{modelEnAlt}
E = \frac{1}{2} \sum_{b=1}^6  s_b^2 +
\frac{1}{2} \sum_{B=1}^{12}  v_B \sigma_B^2 \, ,
\end{align}
where 
\begin{align}
\sigma_B = \uv_{b, \perp} \cdot a_{b^\prime} + \uv_{b^\prime, \perp} \cdot a_{b} \, ,
\end{align} 
and 
\begin{align}
v_B = \frac{v}{|\xv_b|^2 |\xv_{b^\prime}|^2 \sin^2 \theta_B^0} \, .
\end{align} 
Note that this effective bending stiffness is larger than the bare $v$. For our $X_{\text{nt}}$ and $X_{\text{t}}$ lattices, the average of $v_B$ over all 12 angles per unit cell is roughly 100 times larger than $v$. This must be taken into account when comparing results for the GKL with NNN and bending energies, respectively, see below.

\subsubsection{Lattice theory -- compatibility matrix}
The bulk compatibility matrix of our model lattice in $\brm{q}$-space reads
\begin{widetext}
\begin{align}
&\brm{C} (\brm{q}) =
\nonumber\\
& \left(
\begin{array}{cccccc}
a_{1,x} & a_{1,y}  & -a_{1,x} & -a_{1,x} & 0 & 0
\\
0 & 0 & a_{2,x} & a_{2,y} & -a_{2,x} & -a_{2,y}
\\
-a_{3,x} & -a_{3,y}  & 0 & 0 & a_{3,x} & a_{3,y}
\\
-a_{4,x} e^{-i \qv \cdot \Tv_1}& -a_{4,y} e^{-i \qv \cdot \Tv_1}  & a_{4,x} & a_{4,y} & 0 & 0
\\
0 & 0 & -a_{5,x} e^{i \qv \cdot \Tv_3} & -a_{5,x} e^{i \qv \cdot \Tv_3} & a_{5,x} & a_{5,y}  
\\
-a_{6,x} e^{-i \qv \cdot \Tv_2}  & -a_{6,y} e^{-i \qv \cdot \Tv_2} & 0 & 0 & a_{6,x} & a_{6,y} 
\\
a_{1, x}^{\perp3} -a_{3, x}^{\perp1} & a_{1, y}^{\perp3} -a_{3, y}^{\perp1}&  a_{3, x}^{\perp1} &  a_{3, y}^{\perp1} & -a_{1, x}^{\perp3}  & -a_{1, y}^{\perp3} 
\\
-a_{2, x}^{\perp1} & -a_{2, y}^{\perp1}& a_{2, x}^{\perp1} - a_{1, x}^{\perp2} & a_{2, y}^{\perp1} - a_{1, y}^{\perp2} & a_{1, x}^{\perp2} &  a_{1, y}^{\perp2}
\\
a_{2, x}^{\perp3} &a_{2, y}^{\perp3}& -a_{3, x}^{\perp2} & -a_{3, y}^{\perp2}& a_{3, x}^{\perp2}-a_{2, x}^{\perp3} & a_{3, y}^{\perp2}-a_{2, y}^{\perp3}
\\
(a_{1, x}^{\perp6} - a_{6, x}^{\perp1})  e^{-i \qv \cdot \Tv_2} & (a_{1, y}^{\perp6} - a_{6, y}^{\perp1})  e^{-i \qv \cdot \Tv_2} & a_{6, x}^{\perp1}  e^{-i \qv \cdot \Tv_2} & a_{6, y}^{\perp1}  e^{-i \qv \cdot \Tv_2} & - a_{1, x}^{\perp6} & - a_{1, y}^{\perp6}
\\
a_{2, x}^{\perp4}e^{-i \qv \cdot \Tv_1} & a_{2, y}^{\perp4}e^{-i \qv \cdot \Tv_1} & -(a_{4, x}^{\perp2} + a_{2, x}^{\perp4} ) & -(a_{4, y}^{\perp2} + a_{2, y}^{\perp4}) &  a_{4, x}^{\perp2}  &  a_{4, y}^{\perp2} 
\\
a_{5, x}^{\perp3} & a_{5, y}^{\perp3} & a_{3, x}^{\perp5} e^{i \qv \cdot \Tv_3} & a_{3, y}^{\perp5} e^{i \qv \cdot \Tv_3} & -(a_{5, x}^{\perp3} + a_{3, x}^{\perp5}) & -(a_{5, y}^{\perp3} + a_{3, y}^{\perp5})
\\
a_{3, x}^{\perp4} + a_{4, x}^{\perp3} & a_{3, y}^{\perp4} + a_{4, y}^{\perp3}& -a_{3, x}^{\perp4}e^{i \qv \cdot \Tv_1} & -a_{3, y}^{\perp4}e^{i \qv \cdot \Tv_1}& -a_{4, x}^{\perp3} & -a_{4, y}^{\perp3}
\\
-a_{5, x}^{\perp1}e^{i \qv \cdot \Tv_3} &- a_{5, y}^{\perp1}e^{i \qv \cdot \Tv_3} & (a_{5, x}^{\perp1} + a_{1, x}^{\perp5}) e^{i \qv \cdot \Tv_3} &  (a_{5, y}^{\perp1} + a_{1, y}^{\perp5}) e^{i \qv \cdot \Tv_3} & -a_{1, x}^{\perp5} & -a_{1, y}^{\perp5} 
\\
a_{2, x}^{\perp6}e^{-i \qv \cdot \Tv_2} & a_{2, y}^{\perp6}e^{-i \qv \cdot \Tv_2} & -a_{6, x}^{\perp2} & -a_{6, y}^{\perp2} & a_{6, x}^{\perp2} - a_{2, x}^{\perp6} & a_{6, y}^{\perp2} - a_{2, y}^{\perp6}
\\
-(a_{4, x}^{\perp6}+a_{6, x}^{\perp4})e^{-i \qv \cdot \Tv_2} & -(a_{4, y}^{\perp6}+a_{6, y}^{\perp4})e^{-i \qv \cdot \Tv_2} & a_{6, x}^{\perp4}e^{i \qv \cdot \Tv_3} & a_{6, y}^{\perp4}e^{i \qv \cdot \Tv_3} &a_{4, x}^{\perp6}&a_{4, y}^{\perp6}
\\
a_{5, x}^{\perp4}e^{- i \qv \cdot \Tv_2} & a_{5, y}^{\perp4}e^{- i \qv \cdot \Tv_2} & (a_{4, x}^{\perp5} - a_{5, x}^{\perp4} )e^{i \qv \cdot \Tv_3} & (a_{4, y}^{\perp5} - a_{5, y}^{\perp4} )e^{i \qv \cdot \Tv_3} & -a_{4, x}^{\perp5} & -a_{4, y}^{\perp5}
\\
-a_{5, x}^{\perp6}e^{- i \qv \cdot \Tv_2} & -a_{5, y}^{\perp6}e^{- i \qv \cdot \Tv_2} & -a_{6, x}^{\perp5}e^{i \qv \cdot \Tv_3} & -a_{6, y}^{\perp5}e^{i \qv \cdot \Tv_3} & a_{5, x}^{\perp6} + a_{6, x}^{\perp5} & a_{5, y}^{\perp6} + a_{6, y}^{\perp5}
\end{array}
\right) \, ,
\end{align}
\end{widetext}
where $\av_b^{\perp b^\prime} = P_{b^\prime} \av_b$ is the projection of the bond vector of bond $b$ onto the direction perpendicular to bond $b^\prime$. Note that this compatibility matrix is based on the model elastic energy as written in Eq.~(\ref{modelEnAlt}), i.e., the corresponding spring constant matrix has $1$ and the $v_B$ on its diagonal rather than $1$ and the bare $v$. From here on, the lattice and elastic theory calculations proceed exactly as for the NNN GKL. 

\subsubsection{Results}
Having the compatibility matrix for the bending GKL, we can proceed exactly as for the NNN GKL. Inter alia, we can readily contract from it the bulk dynamical matrix of the lattice theory and then calculate the bulk spectrum. We can decompose it into the layer matrixes that provide the foundation for our lattice theory approach for calculating the surface modes. And, we can extract from it the 2 by 2 effective dynamical matrix of the elastic theory.

\begin{figure*}[h]
\begin{minipage}{5.5cm}
	\includegraphics[width=5.3cm]{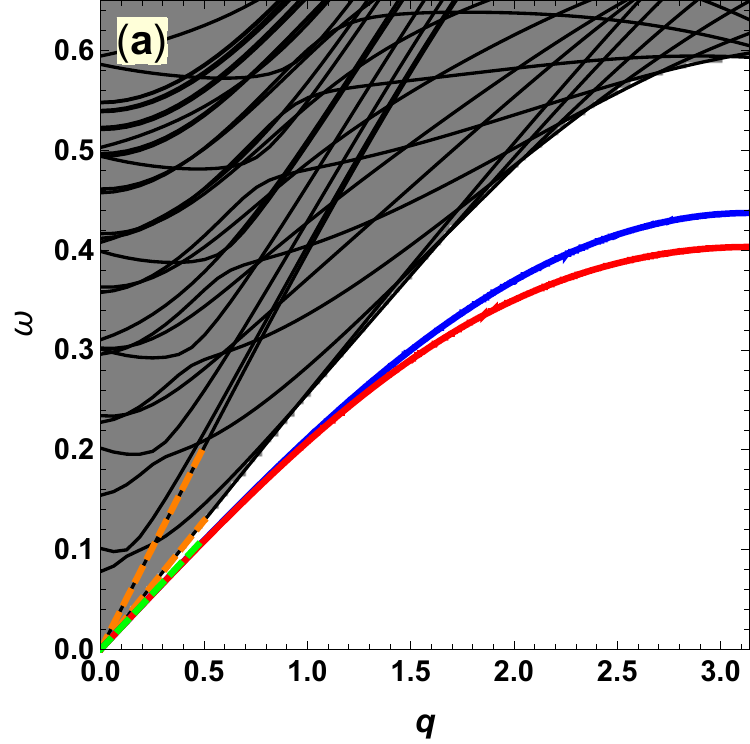}
	\includegraphics[width=5.3cm]{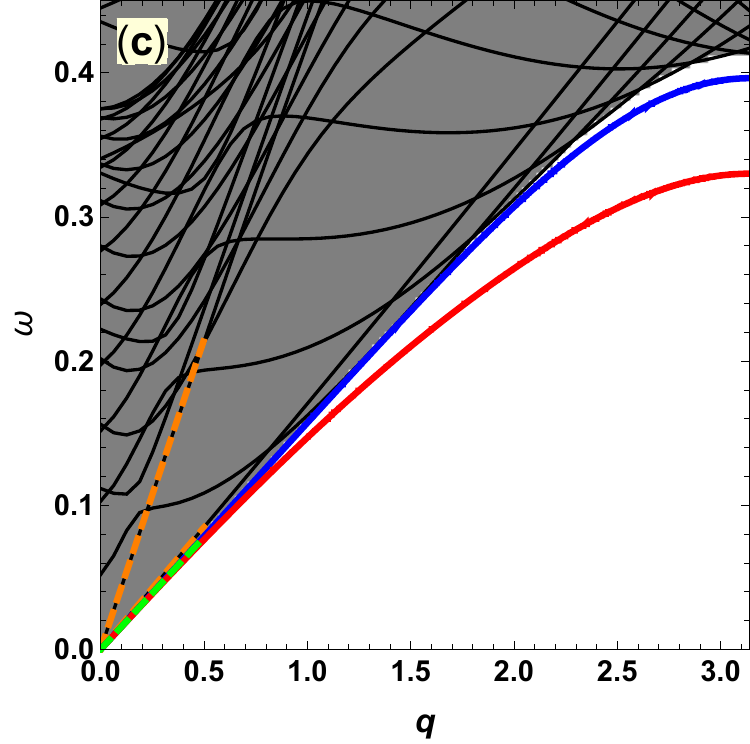}
	\end{minipage}
\begin{minipage}{5.5cm}
	\includegraphics[width=5.3cm]{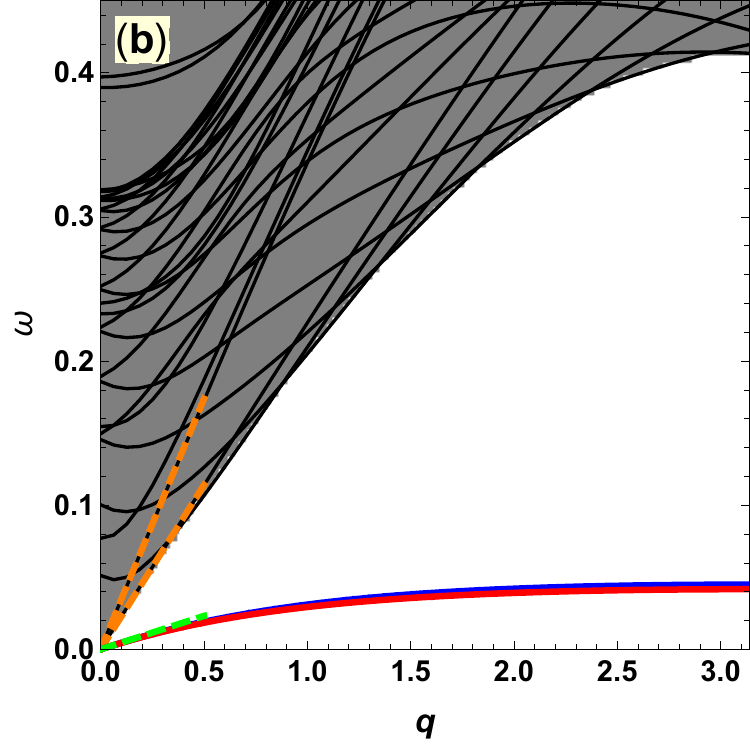}
	\includegraphics[width=5.3cm]{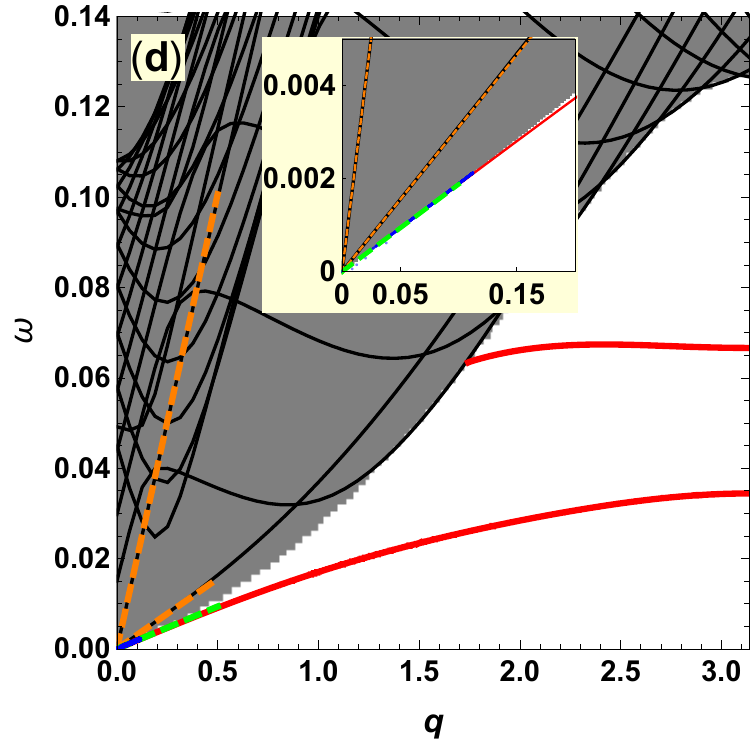}
\end{minipage}
\begin{minipage}{6.5cm}
	 \includegraphics[width=6.0cm]{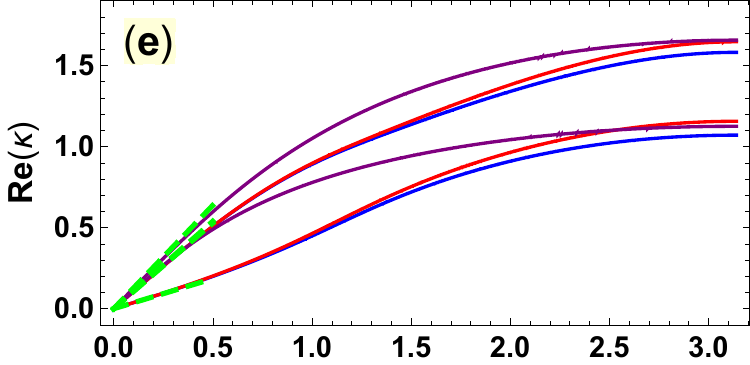}
	\includegraphics[width=6.0cm]{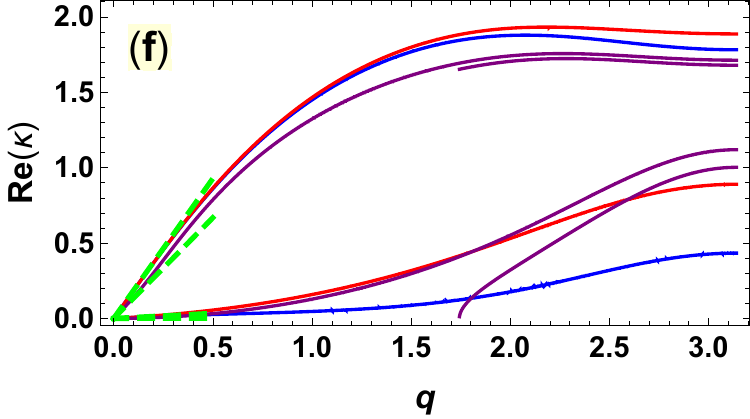}\\
	\vspace{2.5mm}
	\includegraphics[width=6.3cm]{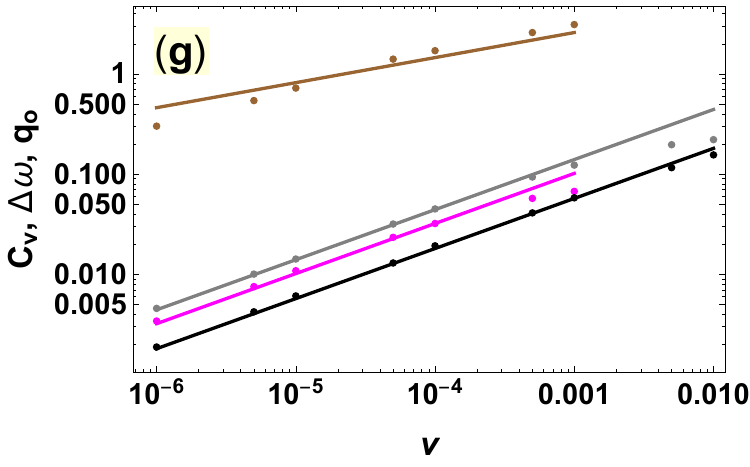}
\end{minipage}
\caption{Low-frequency mode structure for $X_{\text{nt}}$ with (a) $v=0.01$ and (b) $v=0.0001$ and $X_{\text{t}}$ with (c) $v=0.01$ and (d) $v=0.0001$. Inverse penetration depth of the most slowly decaying contributions to the surface modes for (e) $X_{\text{nt}}$ and (f) $X_{\text{t}}$. (g) $c_R$ for $X_{\text{nt}}$ and $c_R$, $\Delta \omega$, and $q_0$ for $X_{\text{t}}$. The lines are power-law fits with $c_R\sim\Delta \omega\sim v^{0.5}$ and  $q_0 \sim v^{0.25}$. The color code is the same as in Fig.~2 of the main paper.}%
	\label{fig:Bendingbands}%
\end{figure*}

Figure~\ref{fig:Bendingbands} compiles our main results for the GKL with bending forces. It shows the low-frequency mode structure for $X_{\text{nt}}$ and $X_{\text{t}}$, the accompanying results for the inverse penetration depths and our our results for the sound velocities $c_R$ for both $X_{\text{nt}}$ and $X_{\text{t}}$, as well as for $X_{\text{t}}$ the vertical gap $\Delta \omega$ at $q=\pi$ between the acoustical surface mode and the lowest optical mode on the bottom surface and the onset $q_0$ of the latter.

As explained above, one should expect that a favorable comparison between the model lattices with NNN and bending forces requires a rescaling of $v$ because it gets, in the model with bending, effectively renormalized to larger values through factors stemming from the rotational invariance of the bending interaction. Comparing Fig.~\ref{fig:Bendingbands} to Fig.~2 of the main paper, we see that the bulk and surface mode frequencies for $X_{\text{nt}}$ are almost identical in both model lattices when $v$ is rescaled in the bending model by a factor of $10^{-1}$. The inverse penetration depths are also very similar in this case. For $X_{\text{t}}$, the results become very similar when $v$ by a factor that is closer to $10^{-2}$. The upshot is that apart from this trivial rescaling, the results for the GKL with NNN and bending forces are very similar, and the signatures of the topological phonons in both are qualitatively the same.

\end{document}